\documentclass[12pt]{article}

\usepackage{graphicx}
\usepackage{amssymb}

\title{Moving Mini-Max -- a new indicator for technical analysis}

\author{Z.~K.~Silagadze 
\vspace*{3mm} \\
Budker Institute of Nuclear Physics and \\ Novosibirsk State
University \\ 630 090, Novosibirsk, Russia}

\date{}

\begin{document}

\maketitle

\begin{abstract}
We propose a new indicator for technical analysis. The indicator emphasizes
maximums and minimums in price series with inherent smoothing and has a 
potential to be useful in both mechanical trading rules and chart pattern
analysis.

\vspace*{2mm}
\end{abstract}

\section{Introduction}
Despite the widespread use of technical analysis in short-term marketing 
strategies, its usefulness is often questioned. According to the efficient 
market hypothesis \cite{1}, no one can ever outperform the market and earn 
excess returns by using only the information that the market already knows.
Therefore, technical analysis, which is based on the price history only, is 
expected to be of the same value for efficient markets as astrology: 
``Technical strategies are usually amusing, often comforting, but of no real 
value'' \cite{2}.

However, the efficient market hypothesis assumes that all market participants 
are rational, while it is a well known fact that human behavior is seldom 
completely rational. Therefore, the idea that one can  try ``to forecast 
future price movements on the assumption that crowd psychology moves between 
panic, fear, and pessimism on one hand and confidence, excessive optimism, 
and greed on the other'' \cite{3} does not seem to be completely hopeless.

At least, ``by the start of the twenty-first century, the intellectual 
dominance of the efficient market hypothesis had become far less universal. 
Many financial economists and statisticians began to believe that stock 
prices are at least partially predictable'' \cite{4}.

Besides, the market efficiency can be significantly distorted at periods of
central bank interventions which allow traders to profit by using even very 
simple technical trading rules at these periods \cite{5,6}.

Anyway the use of technical analysis is widespread among practitioners, 
becoming in fact one of the invisible forces shaping the market. For example,
many successful financial forecasting methods seem to be self-destructive
\cite{4,7}: their initial efficiency disappears  once these methods become 
popular and shift the market to a new equilibrium.

The technical analysis is based on the supposition that asset prices move in
trends and that ``trends in motion tend to remain in motion unless acted upon 
by another force'' (the analogue of the Newton's first law of motion) 
\cite{8}. The financial forces that compel the trend to change are subject of
fundamental analysis \cite{9}. Efficient markets react quickly to various
volatile fundamental factors and to the spread of the corresponding 
information leaving little chance to practitioners of either technical or 
fundamental analysis to beat the market.

However, real markets react with some delay (inertia) to changing financial
conditions \cite{TF} and trends in these transition periods can reveal some 
characteristic behavior determined by human psychology and corresponding
irrational expectations of traders. A skilled analyst can detect these
characteristic features with tools of technical analysis alone (although some
fundamental analysis, of course, might be also helpful and reduce risks). 

Practitioners of the technical analysis often use charting (graphing the 
history of prices over some time period) to identify trends and forecast their 
future behavior \cite{3,8,10,11}. At that peaks and and troughs in the price 
series play important role. Location of such local maximums and minimums is
hampered by short-term noise in the price series and usually some smoothing
procedures are first applied to remove or reduce this noise. 

Below an algorithm for searching of local maximums and minimums is presented.
The algorithm is borrowed from nuclear physics and it enjoys an inherent 
smoothing property. A new indicator of technical analysis, the moving 
mini-max, can be based on this algorithm.

\section{The idea behind the indicator}
The idea behind the proposed algorithm can be traced back to George Ga\-mow's
theory of alpha decay \cite{12}. The alpha particle is trapped in a potential 
well by the nucleus and classically has no chance to escape. However, 
according to quantum mechanics it has non-zero, albeit tiny, probability of 
tunneling through the barrier and thus to escape the nucleus.

\begin{figure}[htb]
\begin{center}
\includegraphics[scale=0.7]{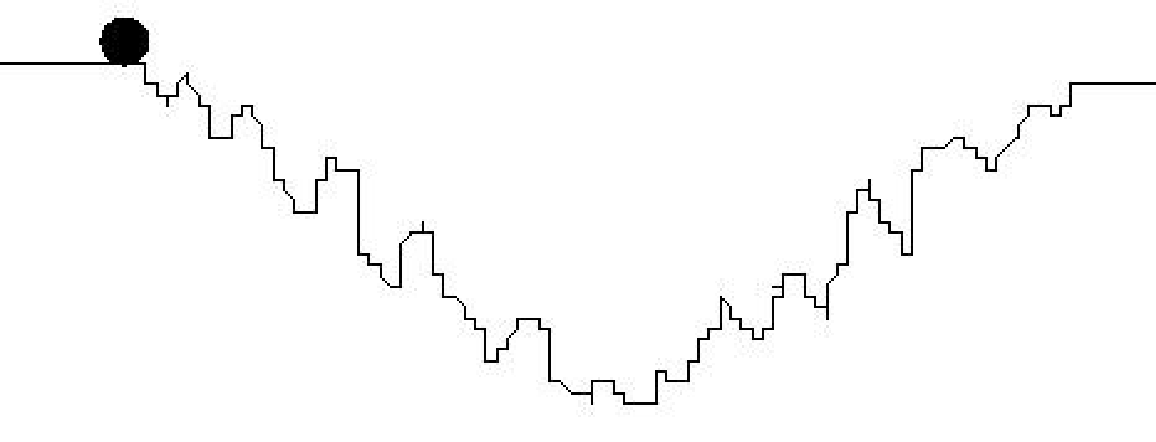}
\end{center}
\caption{A schematic illustration of the idea behind the algorithm: a small
quantum ball can penetrate through narrow barriers and find its way downhill
despite the noise in the potential well shape.}
\label{Fig1}
\end{figure}

Now imagine a small ball placed on the edge of the irregular potential well
(see Fig.\ref{Fig1}). Classical ball will not roll down stopping in front of 
the foremost obstacle. However, if the ball is quantum, so that it can 
penetrate through narrow potential barriers, it will still find its way 
towards  the potential well bottom and oscillate there.

Instead of considering a real quantum-mechanical problem, one can only mimic 
the quantum behavior to reduce the computational complexities. In \cite{13}, 
suitably defined Markov chains were used for this goal. The algorithm that
emerged proved to be useful and statistically robust in $\gamma$-ray 
spectroscopy \cite{14,15}. Two-dimensional generalizations of the algorithm
were also suggested recently \cite{16,17}.

\section{The indicator}
Let $S_i,\; i=1,\ldots, n$ be a price series for some time window. For our
purposes, the moving mini-max of this price series, $u(S)_i$, can be 
considered as a non-linear transformation 
\begin{equation}
u(S)_i=\frac{u_i}{u_1+u_2+\ldots+u_n},
\label{eq1}
\end{equation}
where $u_1=1$ and $u_i,\;i>1$ are defined through the recurrent relations
\begin{equation}
u_i=\frac{P_{i-1,i}}{P_{i,i-1}}\;u_{i-1},\;\;\; i=2,3,\ldots, n.
\label{eq2}
\end{equation}
Evidently, the moving mini-max series satisfies the normalization condition
\begin{equation}
\sum\limits_{i=1}^n u(S)_i =1.
\label{eq3}
\end{equation}
The transition probabilities $P_{ij}$, which just mimic the tunneling 
probabilities of a small quantum ball through narrow barriers of the price 
series, are determined as follows
\begin{equation}
P_{i,i+1}=\frac{Q_{i,i+1}}{Q_{i,i+1}+Q_{i,i-1}},\;\;\;
P_{i,i-1}=\frac{Q_{i,i-1}}{Q_{i,i+1}+Q_{i,i-1}},
\label{eq4}
\end{equation}
with
\begin{equation}
Q_{i,i+1}=\sum\limits_{k=1}^m \exp{\left [\frac{2(S_{i+k}-S_i)}{
S_{i+k}+S_i}\right ]},\;\;\;
Q_{i,i-1}=\sum\limits_{k=1}^m \exp{\left [\frac{2(S_{i-k}-S_i)}{
S_{i-k}+S_i}\right ]}.
\label{eq5}
\end{equation}
Here $m$ is a width of smoothing window. This parameter mimics the (inverse) 
mass of the quantum ball and therefore allows to govern its penetrating 
ability. Besides, it is assumed that $S_{i+k}=S_n$, if $i+k>n$, and
$S_{i-k}=S_1$ if $i-k<1$.

\begin{figure}[htbp]
\begin{center}
\includegraphics[scale=0.7]{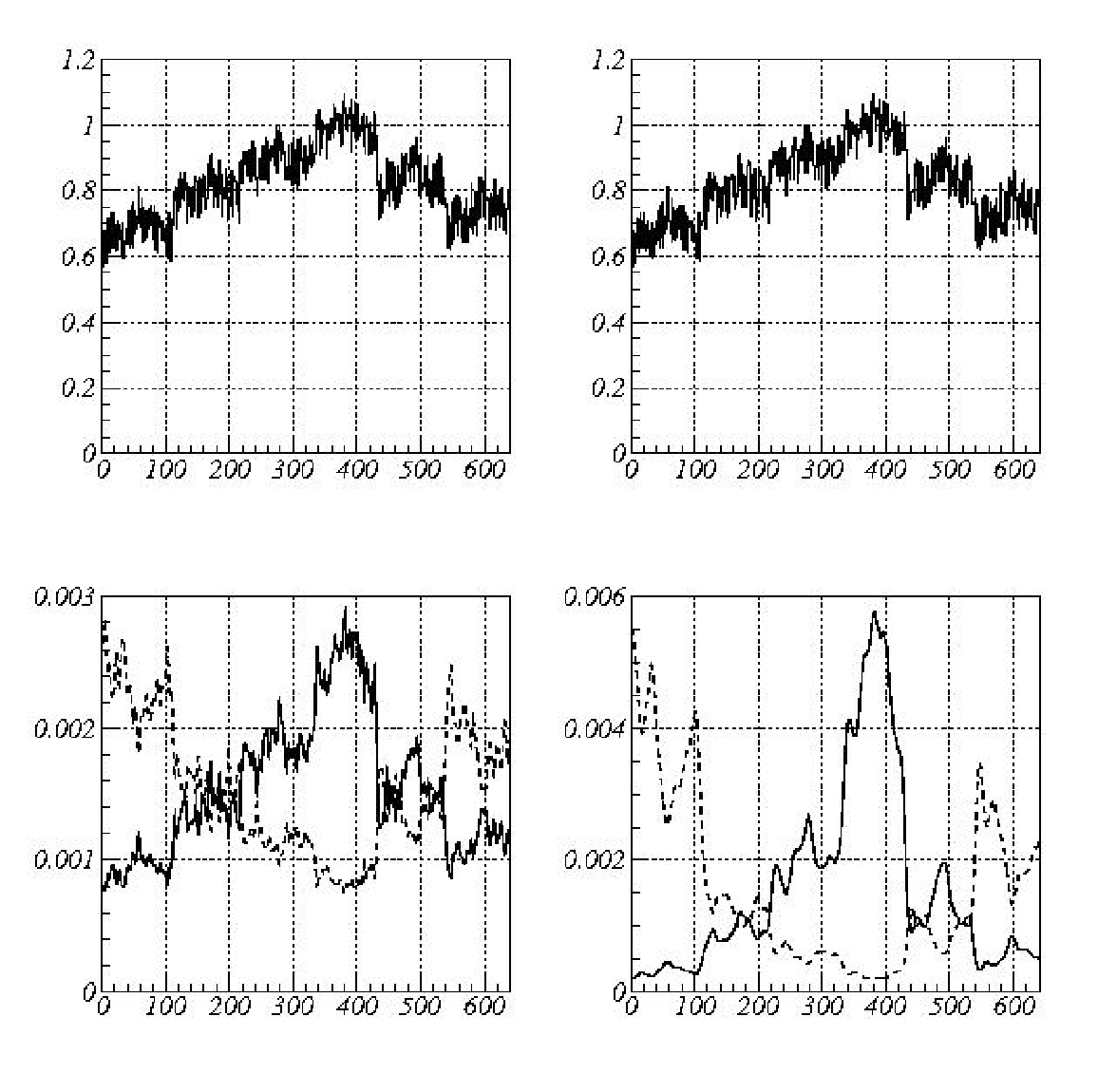}
\end{center}
\caption{A price series $S_i$ (top) and its mini-max (bottom) for the 
smoothing window widths $m=3$ (left) and $m=10$ (right). The solid line 
corresponds to the up mini-max $u(S)_i$ which emphasizes local maximums and 
the dashed line -- to the down mini-max $d(S)_i$ which emphasizes local 
minimums.
}
\label{Fig2}
\end{figure}

The moving mini-max $u(S)_i$ emphasizes local maximums of the primordial 
price series $S_i$ as illustrated by Fig.\ref{Fig2}. Its inherent smoothing
property is also clearly seen in this figure. 

Alternatively, we can  construct the moving mini-max $d(S)_i$ which will 
emphasize local minimums. All what is needed is to change $Q_{i,i\pm 1}$ in 
the above formulas with $Q^\prime_{i,i\pm 1}$ defined as follows
\begin{equation}
Q^\prime_{i,i+1}=\sum\limits_{k=1}^m \exp{\left [-\frac{2(S_{i+k}-S_i)}{
S_{i+k}+S_i}\right ]},\;\;\;
Q^\prime_{i,i-1}=\sum\limits_{k=1}^m \exp{\left [-\frac{2(S_{i-k}-S_i)}{
S_{i-k}+S_i}\right ]}.
\label{eq6}
\end{equation}
That is we change sign to the opposite in all exponents while calculating
the transition probabilities.

\section{Possible applications}
\begin{figure}[htbp]
\begin{center}
\includegraphics[scale=0.55]{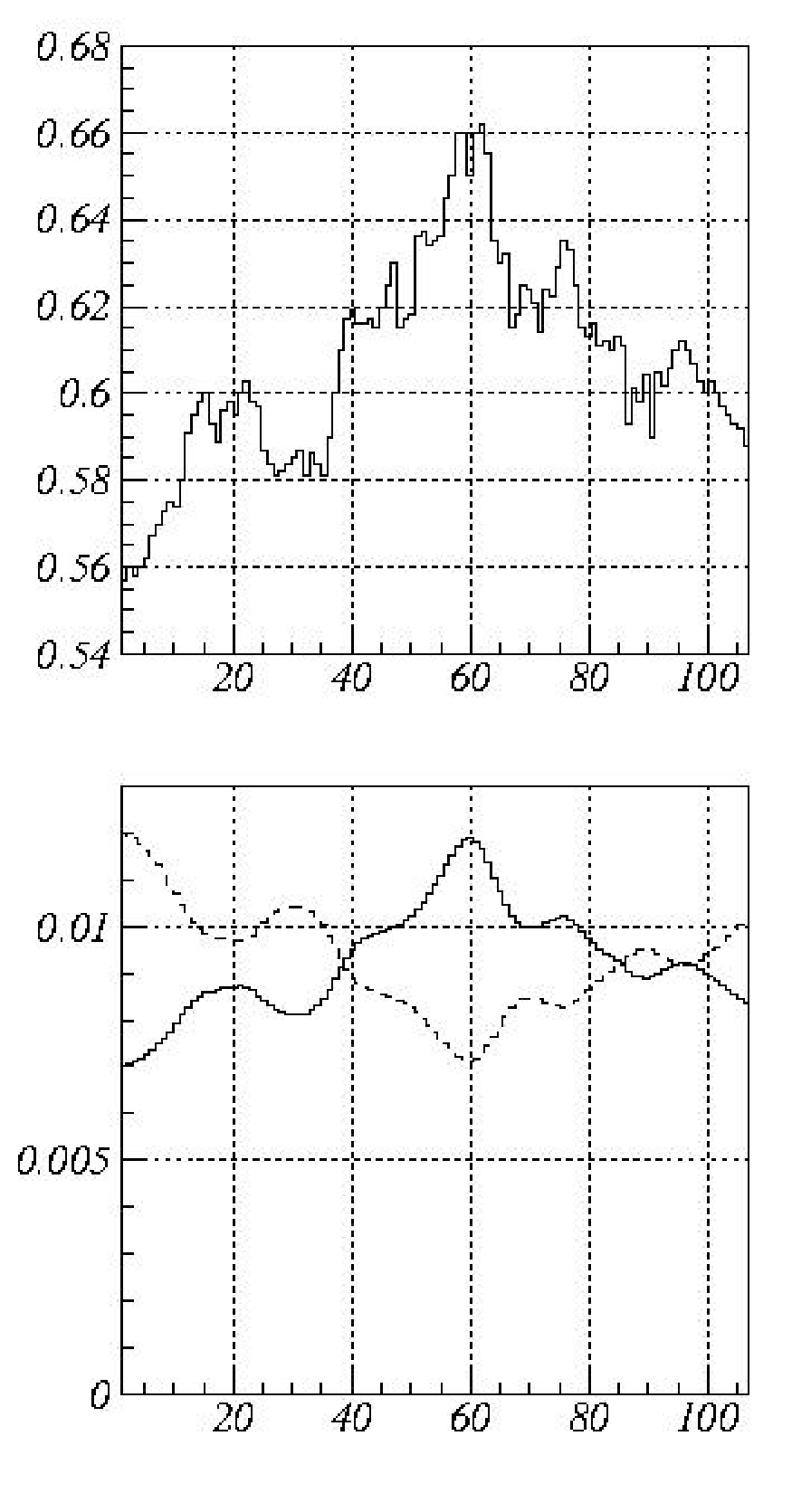}
\end{center}
\caption{A price series $S_i$ (top) that exhibit a head-and-shoulders pattern
 and its mini-max (bottom) for the smoothing window width $m=5$. The solid
 line corresponds to the up mini-max $u(S)_i$ and the dashed line -- to the
 down mini-max $d(S)_i$.
 }
\label{Fig3}
\end{figure}

Do not trying to foresee the imagination of practitioner traders, we indicate
only several possible applications of the new indicator which lay rather
on the surface. 

Resistance and support lines play an important role in technical analysis
\cite{10,11}. To identify lines of resistance and support, traders usually
use some moving average indicator. If the price goes through the local 
maximum and crosses a moving average, we have a resistance line indicating 
the price at which a majority of traders expect that prices will move lower.
A support line happens when the price crosses a moving average after the local
minimum. The support line indicates the price at which a majority of traders 
feel that prices will move higher. The problem is fluctuations of the price
which hampers the identification of both the local extremums and the 
corresponding crossing points with the moving average. The new indicator
can come to the rescue because it naturally suppresses the noise. We can use
$u(S)$ moving mini-max for both the price and its moving average and search
for the crossing points of the corresponding moving mini-maxes to identify
resistance lines. Analogously, $d(S)$ moving mini-maxes can be used to
search for the support lines.

It is widely believed that certain chart patterns can signal either a 
continuation or reversal in a price trend. Maybe the most notorious pattern
of this kind is the head-and-shoulders pattern \cite{18,19}. As the 
identification of this pattern requires to find the extrema of the price 
series, it is evident that the moving mini-max can find its application here.

As an illustration, Fig.\ref{Fig3} shows an alleged head-and-shoulders 
pattern and the corresponding behavior of the  moving mini-max indicators.
Note that $u(S)$ and $d(S)$ indicators form a characteristic spindle like
pattern at the location of the head-and-shoulders. The same behavior is
observed at greater scales in Fig.\ref{Fig2}.

\section{Conclusions}
We hope that the suggested indicator can find its applications in technical 
analysis. ``The classical technical analysis methods of financial indices, 
stocks, futures, \ldots are very puzzling'' \cite{20}. Nevertheless, many
traders find them useful and entertaining. It`s unlikely the new indicator
to disentangle the puzzlement, but we hope it can add some new flavor and 
delight to the occult science of technical analysis. 

\section*{Acknowledgments}
The author thanks V.~Yu.~Koleda who initiated a practical realization of
the suggested indicator and enlightened the author about Forex technical
analysis. The work is supported in part by grants Sci.School-905.2006.2 and
RFBR 06-02-16192-a.

\end{document}